# Divinylglycol, a glycerol-based monomer: valorization, properties and applications

Léa Bonnot[1,2], Christophe Len[3,4], Etienne Grau[1,2] and Henri Cramail*[,1,2]

[1] University of Bordeaux, Laboratoire de Chimie des Polymères Organiques, UMR 5629, Bordeaux INP/ENSCBP, 16 avenue Pey-Berland, F-33607 Pessac Cedex, France.
[2] Centre National de la Recherche Scientifique, Laboratoire de Chimie des Polymères Organiques, UMR 5629, F-33607 Pessac Cedex, France
[3] Sorbonne Universités, Université de Technologie de Compiègne, Centre de Recherches de Royallieu, CS 60319, F-60203 Compiègne cedex, France.
[4] PSL Research University, Chimie ParisTech, CNRS, Institut de Recherche de Chimie Paris, 11 rue Pierre et Marie Curie, F-75231 Paris Cedex 05, France.

*cramail@enscbp.fr

**Abstract**

In the context of the development of bio-refineries, glycerol and its derivatives are co-products of the oleochemistry for which new valorization routes must be found. In this study, the polymerizability of divinylglycol (DVG), a symmetrical C-6 glycerol derivative which bears a vicinal diol and two vinyl functions was studied. The reactivity of the hydroxyl and vinyl functions of DVG through polycondensation and polyaddition reactions was investigated. In a first step, the synthesis of polyesters was carried out by reaction of DVG with various biosourced diesters. In a second route, DVG was polymerized through its vinyl functions by ADMET and thiol-ene addition. Finally, three-dimensional epoxy-amine networks were prepared from a series of diamines and bis-epoxidized DVG, the latter being prepared by oxidation of the DVG double bonds. These different polymerization reactions showed that DVG double bonds were more reactive than the alcohol ones and that a panel of original polymers could be obtained from this bio-sourced synthon.

**Introduction**

Polymers are ubiquitous in our daily lives; the automotive, packaging, health and textile sectors, to name just a few of them, are meaningful. With the depletion of fossil resources and environmental concerns, the search for more sustainable solutions is becoming a necessity. Since the beginning of the 21st century, bio-sourced chemistry has been widely expending with the valorization of biomass, which is an abundant source of carbon structures.[1] In this regard, the concept of a biorefinery was created with the objective of producing feed, food, fuels and molecule platforms for industry.[2] Vegetable oils are the second renewable resource used after ligno-cellulosic biomass.[3] Vegetable oils are inexpensive, non-toxic and have a large potential as precursors of bio-based polymers. These oils are composed of triglycerides, which after reaction with an alcohol, water or a base will produce three molecules of fatty acids (or fatty esters) and a molecule of glycerol. Fatty acids are widely studied as precursors of thermoplastic or thermosetting polymers.[4,5] Glycerol is a simple, non-toxic and versatile molecule that provides access to a wide range of high-value bio-based molecules.[6]

Recently, Len and coll. developed an eco-responsible synthesis of divinylglycol (DVG).[7] Divinylglycol (DVG) also called 1,5-hexadiene-3,4-diol is a symmetrical C-6 monomer which bears a vicinal diol and two vinyl functions. It can be synthesized from mannitol[8–10] or tartaric acid,[11] but this requires protection/deprotection steps and a rather low yield is obtained (20 to 52%). A third route to synthesize DVG is possible by a pinacol coupling of acrolein with a yield of 90% (Scheme 1).[12,13] Acrolein can be synthesized from glycerol but this molecule is toxic; Len and coll. then proposed a direct synthesis of DVG from glycerol.[7]

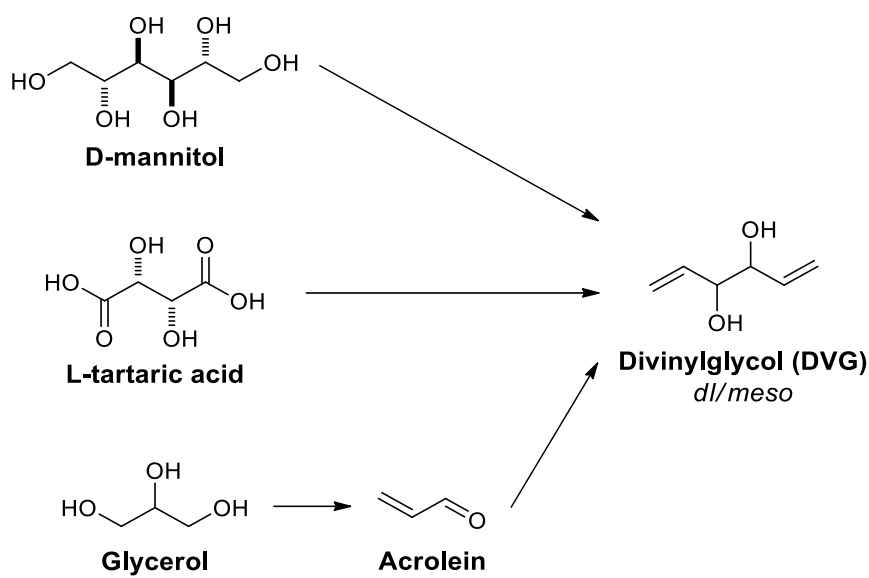

Scheme 1: Synthetic routes to divinylglycol

Divinylglycol is used in the total synthesis of several analogues of natural molecules.[14–16] As polymer precursor, DVG is actually used as cross-linker for thermosets; Noveon® and

Polycarbophil from Lubrizol.[17–19] However, literature data about the reactivity of DVG and the physico-chemical properties of the polymeric materials are very scarce.[20–26]

In this study, the reactivity of DVG with respect to chain-growth and step-growth polymerizations was investigated through the synthesis of linear or cross-linked polymers from the two types of available functions of DVG, the hydroxyl and unsaturation functions, respectively.

I. DVG polymerization via hydroxyl reactivity: example of polyester

The synthesis of DVG-based polyesters by polytransesterification was first investigated. DVG has two secondary alcohol functions, known to be of lower reactivity than that of primary alcohols, in particular in the esterification or transesterification reactions.[27,28] Examples of polyester syntheses from short secondary diols can be find in the literature, in particular 2,3-butanediol which can be obtained by fermentation of glycerol. In the fifties, Watson and coll.[29] have described the synthesis of oligoesters from 2,3-butanediol exhibiting molar masses between 450 and 2,600 g/mol. More recently, bio-sourced polyesters based on 2,5-furandicarboxylic acid and 2,3-butanediol have also been synthesized.[30] Although the formation of rings was demonstrated, polyesters of relatively high molar masses up to 7,000 g/mol were obtained depending on the nature of the catalyst used. Avérous and coll.[31] recently studied the esterification kinetics between 2,3-butanediol and 1,4-butanediol with adipic acid and described the synthesis of (co)polyesters from this mixture of diols.[32]

In this study, the polyesters were prepared in bulk by copolymerization of DVG (mixture of (±) and meso) with dimethylsuccinate (DMSu) or dimethylsebacate (DMSe), as bio-sourced diesters (Scheme 2). Several catalysts were tested such as 1,5,7-triazabicyclo[4.4.0]dec-5-ene (TBD), sodium methanolate (MeONa), titanium butoxide (Ti(OBu)$_4$) and titanium isopropoxide (Ti(O$^i$Pr)$_4$), which are commonly used in polyester synthesis.[33,34]

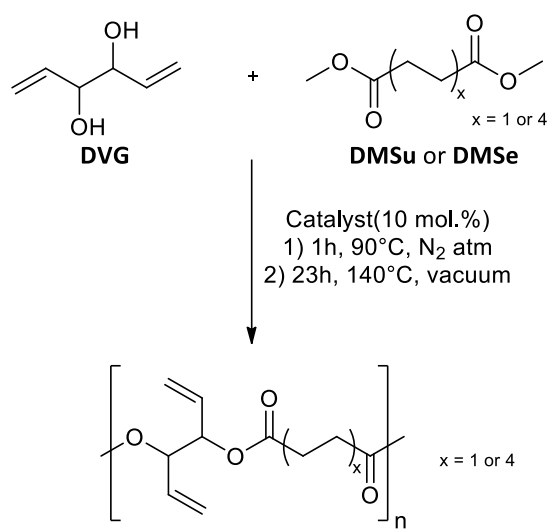

**Scheme 2: Polytransesterification reaction of DVG with DMSu (x = 1) or DMSe (x = 4)**

These experiments show that only oligomers are formed whatever the diester and the catalysts used (Table 1). The apparent molar masses measured by SEC do not exceed 1,840 g/mol for polyesters obtained with DMSe and 1,200 g/mol with DMSu in the presence of TBD, which was the most effective catalyst. Titanium catalysts lead to low conversion of hydroxyl functions of DVG (23 to 50%). These last results can be explained by a possible formation of a complex between the diol and the titanium, which prevents the reaction to take place as already reported by Gau and coll..[35]

**Table 1: Characteristics of polyesters synthesized with DVG and DMSe or DMSu as diester in presence of different catalysts**

| Entry | Diester | Catalyst (10mol.%) | Conversion (%) DVG | Conversion (%) Diester | $M_n$ (a) (g.mol$^{-1}$) | Đ (a) | $T_d$ 5% (°C) | $T_g$ (°C) | $T_f$ (°C) |
|---|---|---|---|---|---|---|---|---|---|
| 1 | DMSe | TBD | 100 | 96 | 950 | 2 | 220 | -39 | - |
| 2 | | MeONa | 100 | 100 | 1,840 | 2 | 260 | -54 | - |
| 3 | | Ti(OBu)$_4$ | 41 | 23 | 500 | 1.2 | 97 | - | -40, 0 |
| 4 | | Ti(O$^i$Pr)$_4$ | 50 | 50 | 700 | 1.6 | 120 | - | -37, -16 |
| 5 | DMSu | TBD | 100 | 90 | 1,200 | 1.2 | 199 | -17 | - |
| 6 | | MeONa | 37 | 84 | 800 | 1.8 | 147 | -47 | - |
| 7 | | Ti(OBu)$_4$ | (b) | (b) | (b) | (b) | 88 | - | - |
| 8 | | Ti(O$^i$Pr)$_4$ | (b) | (b) | (b) | (b) | 88 | - | - |

(a) Measured by SEC in THF, PS calibration, (b) Insoluble polymer

Nevertheless, the various NMR analyses confirm the presence of pendant double bonds along the polyester chains. A chemical shift from 5.81 to 5.74 ppm of the peak corresponding to the protons a' of the double bond of DVG is noticed and protons c' in alpha of the alcohols are shifted from 4 to 5.35 ppm, which confirms the reaction of the alcohol functions (Figure 1).

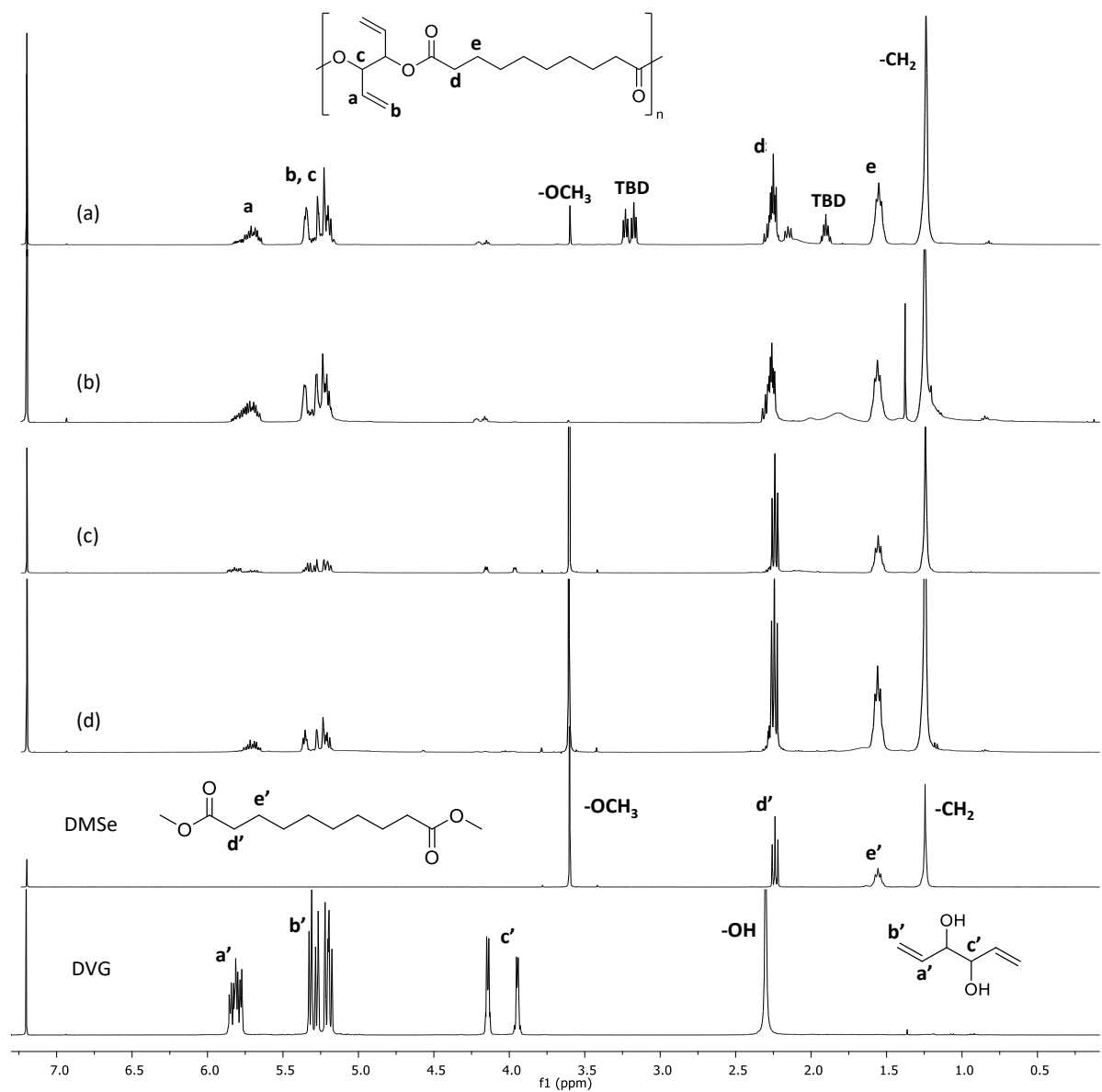

Figure 1: $^1$H NMR in CDCl$_3$ of DVG, DMSe and polyesters synthesized (Entries 1-4, Table 1) ; catalyzed by (a) TBD, (b) MeONa, (c) Ti(OBu)$_4$ and (d) Ti(O$^i$Pr)$_4$

In the following, DVG was further used as a co-monomer (co-diol) with 1,3-propanediol (1,3-PD) or 1,12-dodecanediol (1,12-DD) in the course of polyester synthesis; DMSu or DMSe were kept as diesters and TBD as catalyst (Figure 2). Data are gathered in Table A1 in appendix.

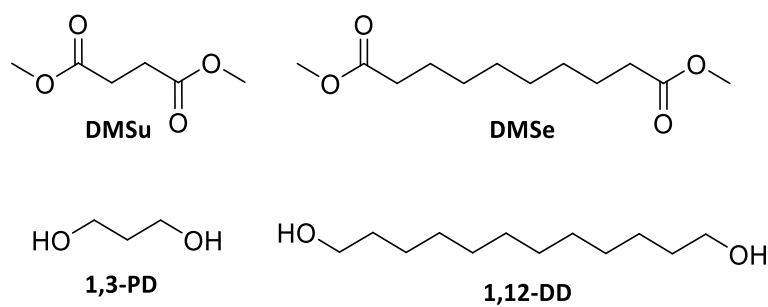

**Figure 2: Co-monomers used for polytransesterification of DVG**

Results obtained are similar with the two diesters used; incorporation of DVG is difficult in both cases and do not exceed 67 and 58% with DMSu and DMSe, respectively. Moreover the incorporation of DVG leads to a decrease of the polymers molecular masses. The copolyesters synthesized with 1,12-DD are semi-crystalline with $T_g$ around room temperature and several melting temperatures, attesting for a heterogeneous composition of the chains. With the short co-diol, 1,3-PD, the copolyesters formed are all amorphous with a $T_g$ between -52 and -40°C.

In conclusion, the synthesis of high molar mass polyesters from DVG revealed impossible. Nevertheless, the advantage of this monomer is the provision of pendant double bonds in the copolymers, allowing a potential post-functionalization that was not addressed in this study.

## II. DVG polymerization *via* double bond reactivity

### a) ADMET polymerization of DVG

Acyclic diene metathesis (ADMET) polymerization follows a step-growth mechanism to produce linear polyalcenamers from $\alpha,\omega$-dienes.[36] ADMET proceeds through a transalkylidenation reaction with the release of ethylene, which can be removed by applying vacuum or a constant flow of an inert gas to obtain high conversions and high molecular weight polymers.[37]

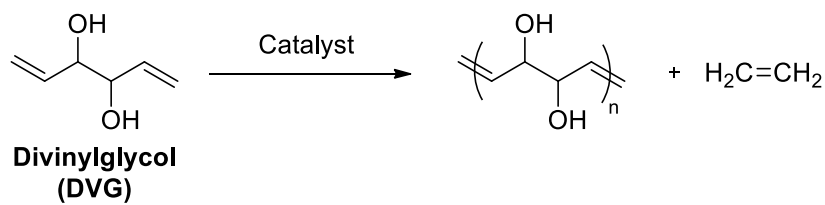

**Scheme 3: ADMET polymerization of DVG**

ADMET polymerization enables the synthesis of polyalcenamers with well-defined architectures and various functions (alcohol, halogen, amine) giving them unique properties.[38,39] The reactivity of DVG and derivatives through ADMET polymerization was thus investigated. Knowing that DVG is a short diene with alcohol functions, the possibility of the negative neighboring group effect (NNGE) with catalysts was analyzed. Wagener and coll.[40] showed that close proximity of functional groups strongly reduces the monomer reactivity, presumably due to complexation of the heteroatom's nonbonded electrons with the metal center. In addition, at higher temperatures, the DVG may be evaporated ($T_b$ = 198°C under 1 atm).

The polymerization of DVG by ADMET (Scheme 3) was tested with five different catalysts, i.e. Schrock catalyst, first and second generation Grubbs catalysts (G1 and G2) and Hoveyda-Grubbs catalysts (HG1 and HG2). The reaction was carried out at 35°C, without solvent and under dynamic vacuum to remove the ethylene formed and shift the equilibrium of the reaction (Table 2).

**Table 2: Characteristics of poly(DVG) synthesized by ADMET polymerization with different catalysts**

| Condition | Catalyst | Quantity of catalyst (mol.%) | $p^{(a)}$ (%) | $M_n^{(b)}$ (g/mol) | $Đ^{(b)}$ | $T_g$ (°C) | $T_d^{5\%}$ (°C) | $X_n$ |
|---|---|---|---|---|---|---|---|---|
| | Schrock | 1 | - | - | - | - | - | - |
| | G1 | 1 | - | - | - | - | - | - |
| Bulk | G2 | 1 | 26 | 780 | 1.2 | -49 | 74 | 1.35 |
| 24h, 35°C | HG1 | 1 | - | - | - | - | - | - |
| Under vacuum | HG2 | 1 | 65 | 300 | 1.5 | -21 | 126 | 2.9 |
| | HG2 | 5 | 68 | 350 | 1.1 | -10 | 77 | 3.1 |
| | HG2 | 0.5 | 75 | 890 | 1.2 | 36 | 122 | 4 |

(a) Conversion of vinyl group calculated by $^1$H NMR, (b) Measured by SEC in DMF, LiBr, PS calibration

As expected, no reaction takes place with the Schrock catalyst which is presumably deactivated by the alcohol functions of DVG. Polymerization also does not occur with G1 and HG1, although these are more robust and polar-resistant than Schrock's catalyst. With the 2nd generation catalysts, an oligomerization reaction is observed. In the case of G2 catalyst, however, the double bond conversion of DVG remains very low (26%), probably because of the low polymerization temperature; indeed, G2 is truly active above 45°C but its activity is 10 times slower that Hoveyda-Grubbs catalyst.[41,42] On the other hand, the reaction is more favorable with HG2, as evidenced by the release of ethylene observable from the addition of the catalyst. Even if this value is still too low, a 65% conversion to poly(DVG) is reached when HG2 is set to 1mol.%. Other tests at HG2 concentrations of 0.5 and 5 mol.% respectively were carried out but, in all cases, the conversion remains of the same order of magnitude and therefore still insufficient to achieve "correct" molar masses.

Other conditions of reaction have been tested such as addition of a polar solvent (THF or water) or an additive (tetrachloro-1,4-benzoquinone or titanium butoxide) to avoid isomerization reaction or to prevent NNGE by hydroxyl screening respectively (Table A2). Better results were obtained when ADMET polymerization was performed in THF with tetrachloro-1,4-benzoquinone (5mol.%). A maximum of 80% of double bond conversion was obtained leading to the formation of oligomers with a degree of polymerization $X_n$ = 5. These last polymerization attempts being carried out in diluted medium, the possibility of forming cycles is favored. MALDI-TOF analysis of the poly(DVG) synthesized in THF in the presence of HG2 as a catalyst demonstrates that no ring was formed and the repeat pattern of P(DVG), 86 g/mol, is clearly visible (Figure 3).

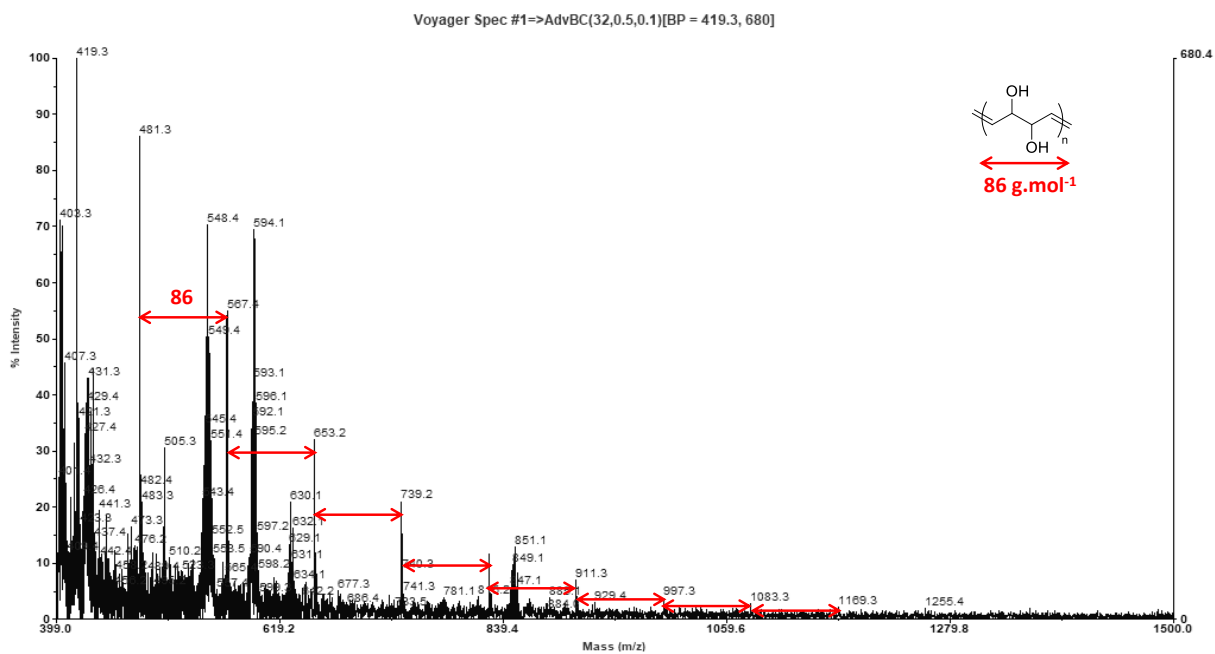

**Figure 3: MALDI-TOF of poly(DVG) synthesized by ADMET polymerization with HG2 as catalyst**

This first study showed that HG2 was the best catalyst for ADMET polymerization of DVG. Even though DVG exhibits an immediate reactivity with HG2, only oligomers were obtained with a maximum conversion of vinyl functions of 80%. Various reactions were tested by adding additives to the medium to suppress the isomerization reaction induced by the catalyst or the NNGE effect of the alcohol functions of DVG. Although the isomerization reaction was suppressed, similar conversions were calculated. The same results were observed by carrying out the reactions in diluted medium in THF. Again DVG revealed quite difficult to polymerize by ADMET.

In the following, DVG was copolymerized with hydrophobic undecyl undecenoate (UndU) (Figure 4) in order to increase the polymer molar mass and also to prepare copolymers exhibiting amphiphilic properties. Indeed polymers obtained by ADMET polymerization of DVG are water soluble. In a first stage, UndU was polymerized by ADMET at 80°C for 24h under dynamic vacuum using HG2 as a catalyst. Thanks to NMR analysis (Figure A1), calculation of the conversion of the vinyl functions of UndU (88%) allowed us determining an $X_n$ = 8.3 ($M_{n\,NMR}$ = 2,788 g/mol) while a value of 1,430 g/mol was measured by SEC in THF (Table 3). This difference between the molar mass values is due to the double bond isomerization reaction caused by the ruthenium catalyst (HG2) as confirmed by MALDI-TOF analysis of P(UndU) on which a distribution of the characteristic peaks of the isomerization at an interval of 14 g/mol are detected (Figure A2).

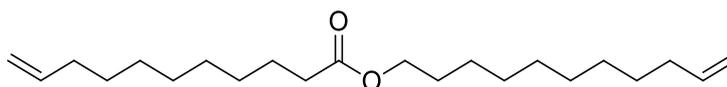

Figure 4: Structure of undecyl undecenoate (UndU)

The poly(UndU) formed was then used for the synthesis of multi-sequenced copolymers by ADMET copolymerization with DVG employed at different ratios (70, 50 and 30mol.%). The $^1$H NMR analysis of the copolymers formed was carried out in CDCl$_3$/DMSO-d6 mixture (50/50 v/v) in order to identify the signal of the two moieties (Figure A3). Low conversions of vinylic functions were obtained in all cases; in the case of DVG/UndU mixture at a ratio 30/70mol.%, the conversion in terminal double bond was high (98%) but the signals corresponding to DVG were not detectable by NMR. For the other two copolymers, DVG signals could be observed and conversions of 24.5 and 36% were calculated for a DVG/UndU ratio of 70/30 and

50/50mol.%, respectively. These conversions only lead to a degree of polymerization ($X_n$) less than 2 (Table 3).

**Table 3: Characteristics of multi-sequenced polymers synthesized by ADMET polymerization from P(UndU) and DVG**

| Product | Ratio DVG/UndU (mol.%) | $p^{(a)}$ (%) | | | $M_n^{(b)}$ (g/mol) | $Đ^{(b)}$ | ΔH (J/g) | $T_f$ (°C) | $T_d^{5\%}$ (°C) | $X_n$ |
|---|---|---|---|---|---|---|---|---|---|---|
| | | DVG | UndU | Total | | | | | | |
| Poly(UndU) | 0/100 | - | - | 88 | 1,430 | 1.6 | 96 | 34 | 227 | 8.3 |
| P(DVG-*b*-UndU) | 70/30 | 32 | 17 | 24.5 | 1,750 | 1.3 | 101 | 33 | 143 | 1.3 |
| | 50/50 | 45 | 26 | 36 | 1,870 | 1.4 | 60 | 34 | 151 | 1.5 |
| | 30/70 | 0 | 98 | 98 | 1,470 | 1.5 | 56 | 33 | 204 | 50 |

(a) Conversion of vinyl group calculated by RMN $^1$H, (b) Measured by SEC in THF, PS calibration

The self-assembly properties in water of the synthesized copolymers composed of hydrophobic (UndU) and hydrophilic (DVG) sequences, were evaluated by nanoprecipitation and analyzed by Dynamic Light Scattering (DLS). The copolymers were dissolved in 0.5 mL of THF at a C = 5 mg/mL and the solution added dropwise (every second) in 4.5 mL of water with a constant stirring of 250 rpm. THF was then evaporated at room temperature and a cloudy solution was obtained visually indicating the presence of dispersed particles (Figure 5). The solution was then filtered on a 0.8 µm filter. DLS analysis was performed 3 times for each sample as reproducibility test.

| P(DVG-*b*-UndU) (70/30) | P(DVG-*b*-UndU) (50/50) | P(DVG-*b*-UndU) (30/70) |
|---|---|---|
| 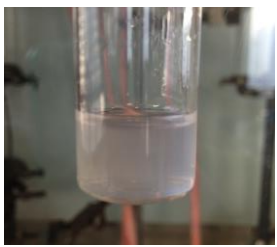 | 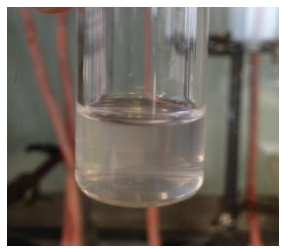 | 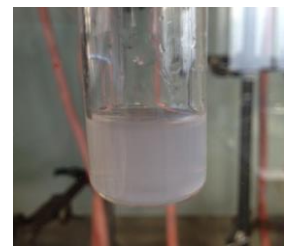 |

*Hydrophobicity* →

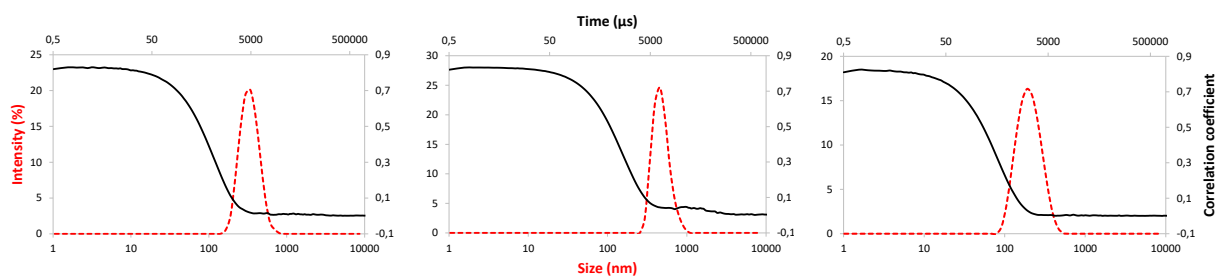

**Figure 5: DLS analysis of multi-block polymers synthesized from P(UndU) and DVG**

These analyses confirm that objects of the order of 200 to 500 nm are formed with polydispersity, PDI, of the order of 0.1 (Figure 5). These dispersions are stable for several months, except for the 50/50 copolymer that contains the largest objects. One should note that the P(UndU) do not lead to stable particle by this process indicating that the 30/70 copolymers contains a small fraction of DVG units.

In this part, multi-sequenced copolymers from DVG and UndU could be synthesized; DVG again shows a low reactivity by ADMET. However, these copolymers exhibit amphiphilic behaviors as confirmed by NMR and DLS analyses.

### b) Polymerization of DVG by thiol-ene reaction

A series of aliphatic dithiols were tested in the course of thiol-ene addition to DVG following two procedures. The monomers were used at the 1:1 stoichiometry and the reaction conducted either at 80°C for 24 h or under UV (65 mW/cm$^2$ at 365 nm) for 15 min; AIBN and Irgacure 2659 were used as thermo- and photo-initiator, respectively (Figure 6).

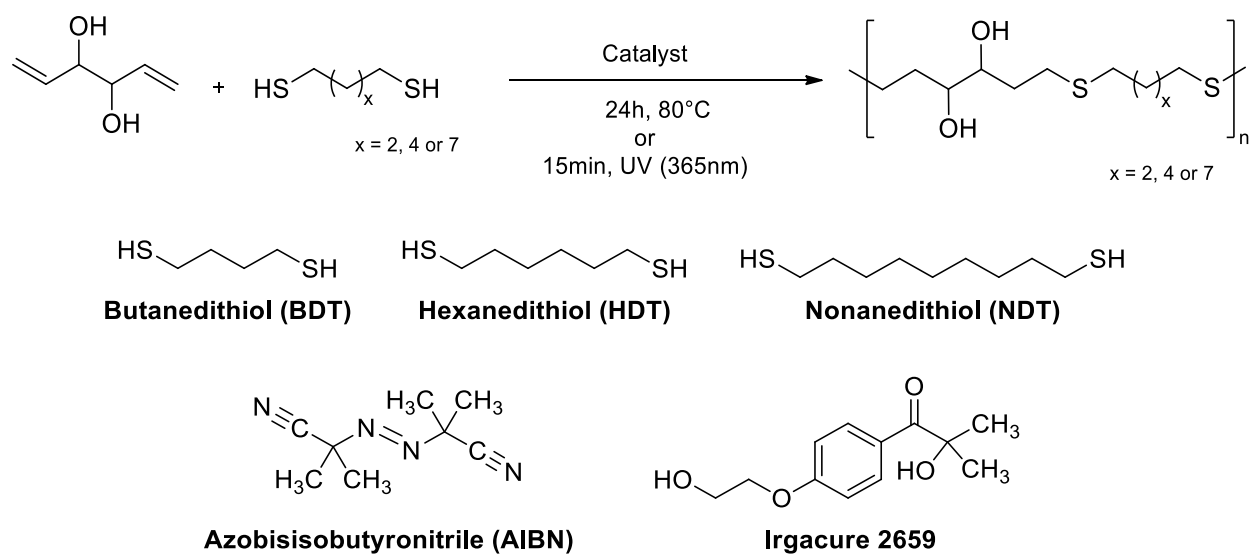

**Figure 6: Thiol-ene reaction of DVG and structure of dithiols and initiators used**

Results obtained show that the photochemical initiation (procedure b) is more efficient than the thermal initiation (procedure a) (Table 4). Indeed, by photopolymerization, polymers of high molar masses (6,000 g/mol) were obtained in short time (15 min). In the case of polymerizations carried out under UV, no kinetic monitoring was carried out but the polymers were analyzed by $^1$H NMR after 15 min of irradiation; conversions of double bonds from 93 to 96% could be calculated. The various reactions tested indicate that the chosen aliphatic dithiols have different reactivities with DVG. The best system is with HDT under photochemical activation (Table 4, entry 4); a total conversion is obtained in few hours.

**Table 4: Characteristics of polymers synthesized by thiol-ene reaction with DVG**

| Entry | Dithiol | Procedure | p (%) | $X_n$[$M_n$ RMN] | $M_n^{(c)}$ (g.mol$^{-1}$) | $Đ^{(c)}$ | $T_d^{5\%}$ (°C) | $T_g$ (°C) | $T_f$ (°C) |
|---|---|---|---|---|---|---|---|---|---|
| 1 | BDT | (a) | 93 | 14.3[3375] | 2,900 | 1.7 | 262 | -26 | - |
| 2 | | (b) | 93 | 14.3[3375] | 7,400 | 1.5 | 277 | -22 | - |
| 3 | HDT | (a) | 93 | 14.3[3775] | 4,000 | 2 | 232 | -14 | 73 |
| 4 | | (b) | 96 | 33[8712] | 6,600 | 1.6 | 279 | -25 | 83 |
| 5 | NDT | (a) | 83 | 5.9[1800] | 3,700 | 1.7 | 267 | 35 | 89 |
| 6 | | (b) | 94 | 16.7[5100] | 6,000 | 1.5 | 275 | 35 | 90 |

(a) AIBN (5mol.%), 24h à 80°C, (b) Irgacure (0,05mol.%), 15min under UV (365 nm), (c) Measured by SEC DMF, LiBr, PS calibration.

NMR analysis also shows that the alcohol functions of DVG (characteristic signal at 4.4 ppm) are still present (Figure 7). Moreover, whatever the initiation procedure, NMR spectra of the polymers formed reveal the presence of multiplets at 0.9 ppm which are characteristic of methyl groups (-CH$_3$). This phenomenon indicates the possible addition of the radical on the most substituted carbon of the DVG double bond whereas anti-Markovnikov addition is generally expected.

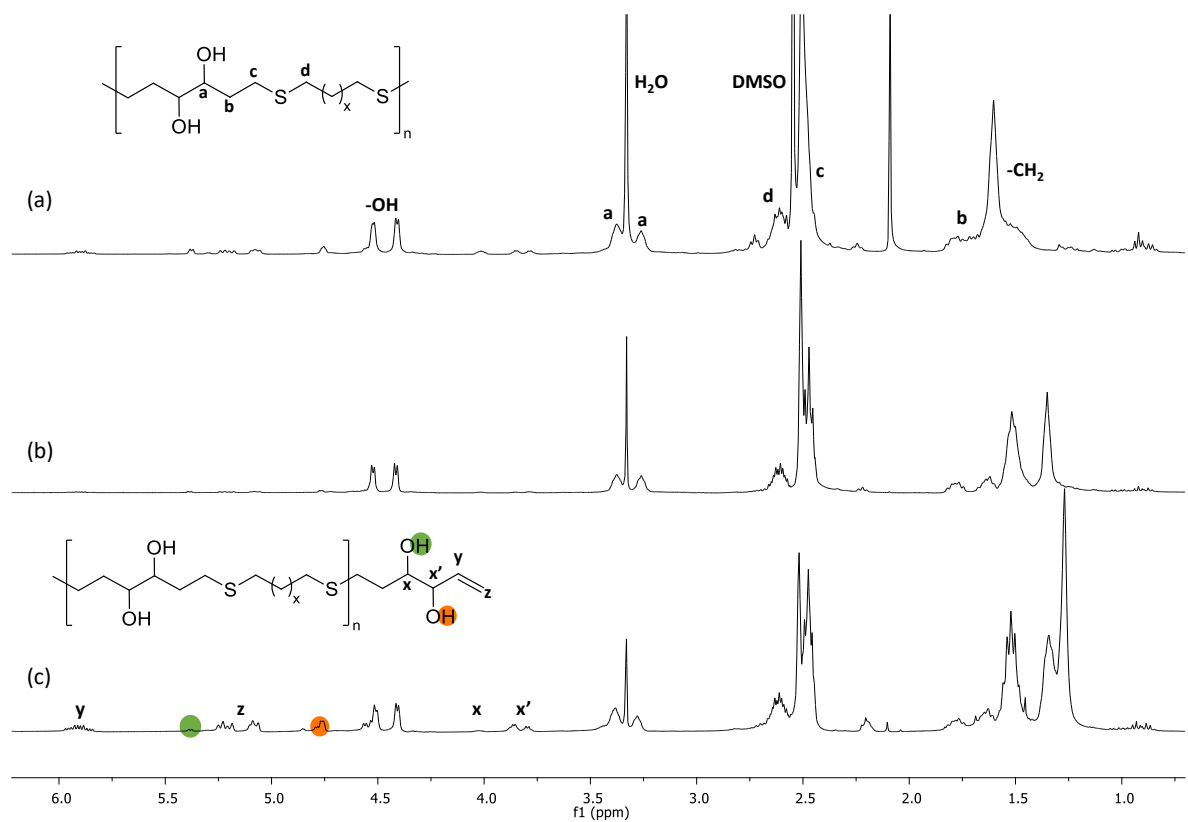

**Figure 7: $^1$H NMR in DMSO of polymers synthesized with DVG and a) BTD, b) HDT or c) NDT by photopolymerization**

In conclusion of this section, DVG exhibits a good reactivity towards thiol-ene reactions through thermal or photochemical initiation. The light-curing process enables to synthesize polymers in a simple and rapid manner with conversions of DVG double bonds higher than 93%, in 15 min. Although DVG is hydrophilic and provides hydroxyl functions, the polymers synthesized by thiol-ene addition from DVG and linear dithiols revealed insoluble in water.

Thiol-ene reaction provides access to DVG-based polymers with relatively high molar masses indicating that vinylic functions of DVG display a good reactivity. The synthesis of cross-linked materials by thiol-ene reaction was thus carried out in the presence of a symmetric tetrathiol

(PTM), which have the advantage to be not odorous, and Irgacure 2659 as a photo-initiator (Figure 8). The thiol-ene reaction was conducted by mixing DVG and thiol monomer at different ratios. The initiator was then added (0.5 mol.%) and the mixture poured into a Teflon mold of 7.2×5.1cm and then cross-linked under UV (365 nm) and a power of 65 mW/cm² for 15 minutes to form films 2 mm thick.

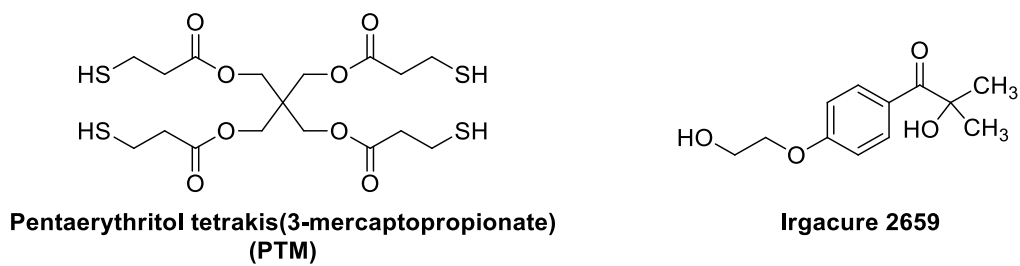

Pentaerythritol tetrakis(3-mercaptopropionate) (PTM)   Irgacure 2659

**Figure 8: Structure of tetrathiol and initiator used for the synthesis of DVG-based thermosets**

Cross-linked materials by thiol-ene reaction were thus synthesized from DVG and tetrafunctional PTM. The thiol/double bond ratio was varied from 50% to 100% in order to measure the impact of the amount of PTM used on the cross-linking reaction and the properties of the networks obtained (Table 5).

**Table 5: Properties of networks synthesized by thiol-ene with DVG and PTM**

| PTM (%) | Swelling THF[H$_2$O] (%) | Soluble part (%) | $T_d^{5\%}$ (°C) | $T_g$ (°C) | $E'_{T\delta+30°C}$ (MPa) | ρ (mol/dm³) | Strain (MPa)[a] | Stress (%)[a] | Modulus (MPa)[a] |
|---|---|---|---|---|---|---|---|---|---|
| 50 | 290[10] | 39[15] | 120 | -33 | 0,15 | 0,057 | 0,1 | 99 ±7.8 | 0.13 |
| 70 | 191[7] | 14[6] | 170 | -22 | 1,8 | 0.67 | 0.6 ± 0.1 | 35.8 ±1.4 | 1.96 ±0.3 |
| 90 | 163[6] | 14[3] | 285 | -17 | 2,5 | 0.93 | 0.6 | 28 ±3.8 | 2.2 ±0.3 |
| 100 | 189[10] | 8[3] | 290 | -16 | 3,5 | 1.3 | 1.2 ± 0.2 | 33.6 ±4.2 | 4.1 ±0.6 |

(a) Average values

Swelling tests in THF were carried out for 24 hours. The swelling ratios vary from 189% to 290% in agreement with the crosslinking density, ρ. Similarly, the soluble fraction decreases with the percentage of PTM used, from 39% to 8%. These experiments confirm that a high content of tetrathiol (PTM) is required to obtain the formation of a 3D-network. Since DVG brings some hydrophilicity through its OH groups, swelling tests in water were also carried out; the swelling ratios, around 10%, are lower in water in comparison to THF, but follow the same trend according to the PTM content. These results were confirmed by FTIR analysis of the films obtained (Figure A4); a characteristic signal of the double bond of DVG is observable at 3080 and 1640 cm$^{-1}$ and decrease with higher PTM concentrations demonstrating the almost complete crosslinking of the polymers.

As a conclusion of this part, 3D networks based on DVG and PTM were prepared. The latter have different mechanical properties whose properties vary with the degree of crosslinking. It was found also possible to vary the hydrophilicity of these materials by varying the thiol used.

### III. DVG-derivate polymerization: example of epoxy-amine network

While DVG direct polymerization remains of low efficiency either using double bond or hydroxyl reactivity, the derivatization of DVG to more reactive function was investigated. At first, DVG was derivatized by oxidation of the double bonds in order to obtain a bis-epoxidized precursor (DAG). DVG was oxidized in dichloromethane during 24h in the presence of MCPBA in excess (Scheme 4). A maximum yield of 50% was obtained with incomplete conversion of DVG. DAG was then reacted with a series of diamines to prepare epoxy networks.

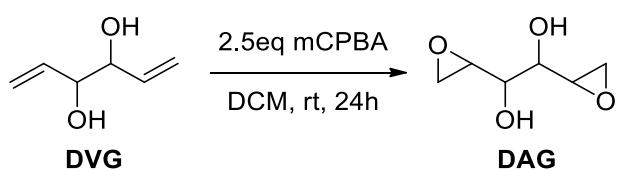

**Scheme 4 : Epoxidation reaction of DVG with mCPBA**

Syntheses with short alkyl diamine having primary or secondary amino groups (ethylene diamine, propylene diamine, isophorone diamine, diethylene triamine) lead to 'burned' materials; indeed, enthalpies of reaction values measured by DSC are quite high (422 to 647 J.g$^{-1}$) and the reaction starts at room temperature (Table A3). These preliminary investigations clearly demonstrated the high reactivity of DAG. To avoid this issue, diamines with long alkyl chains were selected, such as *Priamine® 1075* which is a bio-sourced derivative from fatty acid and *Jeffamine ED600* and *ED900* which are oligo-ethers with molar masses equal to 678.8 and 972 g/mol, respectively. The curing reaction was followed by DSC. Enthalpy of reaction (ΔH), reaction start temperature (T$_{on\ set}$) and glass transition temperature (T$_g$) were determined by

DSC (Table 6). These analyzes reveal relatively low glass transition temperature values in agreement with the $T_g$ of *Jeffamine* precursors, -39 and -45°C respectively. A $T_g$ of 26°C was measured for the networks formed with *Priamine® 1075* which brings more rigidity. Formulation of networks was carried out from EEW (epoxy equivalent weight, g/eq.) value of DAG and from AHEW (amine hydrogen equivalent weight, g/eq.) value of the diamine. The curing of epoxy/amine formulation is theoretically performed with a ratio of 1 mole of epoxy group with 1 mole of active hydrogen from amine.

Table 6: Characteristics of the reactivity of diamines with DAG

| Diamine | f | AHEW (g.eq$^{-1}$) | ΔH (J.g$^{-1}$) | $T_{on\ set}$ (°C) | $T_g$ (°C) |
|---|---|---|---|---|---|
| **Priamine® *1075*** | 4 | 136.17 | 433 | 36 | 26 |
| **Jeffamine ED600** | 4 | 169.7 | 353 | 62 | -39 |
| **Jeffamine ED900** | 4 | 243 | 263 | 71 | -45 |

Networks obtained with these diamines display different properties. With *Priamine® 1075*, the curing is performed at room temperature during 24h. Swelling tests were made in THF and water (Table 7).

Table 7: Characteristics of DAG/ *Priamine® 1075* network

| Diamine | Swelling (%) THF [H$_2$O] | Soluble part (%) | $T_d^{5\%}$ (%) | $T_\alpha$ ($T_g$) (°C) | $E'_{T\alpha+30°C}$ (MPa) | ρ (mol/dm$^3$) |
|---|---|---|---|---|---|---|
| *Priamine® 1075* | 180 [20] | 34 [2] | 287 | 22(19) | 6.2 | 1.6 |

The network swelling was higher in THF (180%) and revealed a fairly large soluble fraction of 34%, demonstrating that the crosslinking was not complete. In water, due to its chemical structure, DAG provides hydroxyl functions and the opening of the epoxide ring generates new ones, which explains a certain affinity of this material with water. The proportion of soluble fraction remains nevertheless very low at 2%. Mechanical measurements by DMA

(Dynamic mechanical analysis) allowed us estimating a crosslink density of 1.6 mol/dm³ and measuring a phase transition, Tα around 20°C.

With *Jeffamines*, networks were cured for 1h at 120°C and 2h at 140°C, the complete polymerization was confirmed by DSC (Table A4). The network prepared from *Jeffamine ED600* was similar to a viscous fluid while an aspect close to an elastomeric gum was obtained from *Jeffamine ED900*. Sticky materials were obtained so tack properties of these networks were measured by DMA (Figure A5); tack is measured as the force required to separate adhesive from a probe at the interface shortly after they have been brought into contact under a defined duration at a specific temperature (here, 10s at room temperature).

Diglycidyl ether of bisphenol-A (DGEBA) and isophorone diamine (IPDA) are classical components of epoxy-amine networks. One of the major issues with this type of material is the impossibility of degrading them in 'soft' conditions. With this purpose, DAG was incorporated into the DGEBA/IPDA network at different percentages (1, 5 and 10 wt.%) in order to provide 1,2-diol units within the network and then evaluate if the latter could be cleaved after treatment with a strong acid such as periodic acid (Table 8).

Table 8: Properties of DGEBA / IPDA materials with addition of 1, 5 and 10wt.% of DAG

| DAG (w.%) | $T_d^{5\%}$ (%) | $T_\alpha (T_g)$ (°C) | $\rho$ (mol/dm³) | Strain (MPa)[a] | Stress (%)[a] | Young modulus (MPa)[a] | Swelling THF [$H_2O$] | Soluble part (%) |
|---|---|---|---|---|---|---|---|---|
| 0 | 352 | 148(137) | 10.6 | 74 ±11 | 6.3 ±1.1 | 1337 ±102 | 3 [1] | <1 |
| 1 | 341 | 135(122) | 5.9 | 57 ±6.2 | 3.4 ±0.5 | 1971 ±165 | 4 [2] | <1 |
| 5 | 342 | 151(129) | 6.2 | 40 ±14 | 3.4 ±1.3 | 1522 ±17 | 4 [2] | <1 |
| 10 | 336 | 138, 157 | 6.6 | 56 ±12 | 3.9 ±0.8 | 2105 ±150 | 3 [2] | <1 |

(a) *Average values*

This first study demonstrates that DAG could be used as a comonomer in DGEBA/IPDA mixtures without detrimental of the final thermomechanical properties. When DAG was used at 1 and 5wt.%, it is perfectly mixed within the DGEBA/IPDA system and the network so-formed remains homogeneous. However, at 10wt.%, a phase segregation is observed, as attested by the DMA analysis (Figure A6) that reveals the presence of two Tα values and by

DSC analysis, $T_g$ was not detected (Table 8) . IPDA has a similar reactivity with DGEBA and DAG (ΔH = 420 J/g) but the polymerization starts at room temperature with DAG instead of 88°C with DGEBA. Swelling tests in THF and water show that the level of soluble fraction is negligible (less than 1% in both cases), demonstrating that the networks so-formed are therefore completely cross-linked (Table 8).

As already stated, the advantage of incorporating DAG in the DGEBA/IPDA network is to provide 1,2-diol units in the final network that can be cleaved. After 24h of reaction with periodic acid (1eq/DAG) in THF, the remaining polymer was oven-dried and weighed. The mass measured was nearly equal to the initial one indicating that the cleavage did not occur. However, the analysis by SEC of the soluble fraction revealed the presence of oligomers with molar masses between 700 and 6,000 g/mol (RI detection), also higher molar masses are detected by UV which could indicate the presence of aromatic rings resulting from DGEBA unit. These same samples were analyzed by FTIR; a signal at 1700 cm$^{-1}$ is identifiable and may correspond to the C=O elongation of the ketones or aldehydes formed after cleavage of the 1,2-diol moiety (Figure A7). This preliminary test gives some insights of polymer degradation and will be further investigated.

**Conclusion**

In conclusion, this study aims at investigating the polymerizability of DVG in various polymerization reactions for the development of original polymer materials. This work allowed us a better understanding of the reactivity of DVG and the access to potentially valuable oligomers and networks. However, this work also reveals the very low reactivity of DVG due to its very specific chemical structure, i.e the close vicinity of hydroxyl and unsaturation functions. Thiol-ene reaction was found the most realistic and efficient reaction to prepare linear polymer and cross-linked materials from DVG. These materials have pendant hydroxyl functions which can be post-functionalized in order to bring new properties and can be potentially degraded by cleavage of 1,2 diol moieties in acidic conditions.

**Acknowledgements**

This work was performed, in partnership with the SAS PIVERT, within the frame of the French Institute for the Energy Transition (Institut pour la Transition Energétique - ITE) P.I.V.E.R.T. (www.institut-pivert.com) selected as an Investment for the Future ("Investissements d'Avenir"). This work was supported, as part of the Investments for the Future, by the French Government under the reference ANR-001-01.

Appendix

**Table A1: Characteristics of copolyesters synthesized from DVG and a co-diol (1,3-PD or 1,12-DD) with a diester (DMSu or DMSe), in the presence of TBD as a catalyst (2mol.%)**

| Diester [Diol] | Ratio DVG/Diol (mol.%) | Conversion (%) | | | Final ratio DVG/Diol (mol.%) | $M_n$ [a] (g/mol) | $Đ$ [a] | $T_d{}^{5\%}$ (°C) | $T_g$ (°C) | $T_f$ (°C) |
|---|---|---|---|---|---|---|---|---|---|---|
| | | DVG | Diol | Diester | | | | | | |
| DMSu [1,3-PD] | 0/100 | - | 98 | 88 | - | 2,000 | 1,4 | 140 | -48 | - |
| | 30/70 | 58 | 100 | 88 | 20/80 | 1,200 | 1,3 | 181 | -40 | - |
| | 50/50 | 59 | 96 | 91 | 43/57 | 620 | 1,3 | 137 | -52 | - |
| | 70/30 | 74 | 95 | 64 | 67/33 | 650 | 1,3 | 132 | -48 | - |
| DMSu [1,12-DD] | 0/100 | - | 98 | 100 | - | 3,600 | 1,9 | 281 | - | 75 |
| | 30/70 | 100 | 75 | 89 | 12/88 | 2,800 | 1,7 | 109 | 26 | 67 |
| | 50/50 | 21 | 92 | 77 | 49/51 | 970 | 1,3 | 78 | 17 | 55, 59 |
| | 70/30 | 40 | 99 | 72 | 59/41 | 430 | 1,6 | 95 | 10 | -14, 35, 43 |
| DMSe [1,3-PD] | 0/100 | - | 100 | 80 | - | 1,900 | 1,3 | 166 | nd | nd |
| | 30/70 | 50 | 95 | 72 | 21/79 | 1,000 | 1,2 | 124 | -44 | 15, 33 |
| | 50/50 | 40 | 95 | 43 | 29/71 | 630 | 1,5 | 132 | -47 | 13, 29 |
| | 70/30 | 77 | 99 | 56 | 50/50 | 630 | 1,4 | 124 | -48 | -13, -9, -4 |
| DMSe [1,12-DD] | 0/100 | - | 100 | 100 | - | 6,700 | 1,8 | 254 | 50 | 80 |
| | 30/70 | nd | 100 | 95 | 4/96 | 3,300[b] | 2[b] | 269 | nd | 76 |
| | 50/50 | 100 | 100 | 53 | 32/68 | 1,450[b] | 1,8[b] | 150 | nd | 31, 61, 66 |
| | 70/30 | 100 | 100 | 63 | 58/42 | 1,000[b] | 1,7[b] | 158 | nd | 27, 47 |

(a) Measured by SEC in THF, PS calibration

**Table A2: ADMET polymerization of DVG with HG2 as a catalyst in the presence of a solvent or an additive**

| Conditions | HG2 quantity (mol%) | Additive (5mol%) | $p^{(a)}$ (%) | $M_n^{(b)}$ (g/mol) | $Đ^{(b)}$ | $T_g$ (°C) | $T_d^{5\%}$ (°C) | $X_n$ |
|---|---|---|---|---|---|---|---|---|
| Bulk 24h, 35°C | 1 | Chloranil | 74 | 890 | 1,1 | 51 | 149 | 4 |
|  |  | Ti(OBu)$_4$ | 64,5 | 715 | 1,2 | -32 | 81 | 2,8 |
| H$_2$O 48h, 35°C | 1 | - | 63,5 | 690 | 1,1 | 11 | 120 | 2,7 |
| THF 24h, 35°C | 0,5 | - | 76 | 840 | 1,2 | 18 | 90 | 4,2 |
|  | 1 | - | 75 | 900 | 1,2 | 33 | 97 | 4 |
|  | 1 | Chloranil | 80 | 1,000 | 1,1 | 45 | 142 | 5 |
|  | 5 | Ti(OBu)$_4$ | 78 | 940 | 1,3 | 44 | 116 | 4,6 |

(a) Calculated by RMN $^1$H, (b) Measured by SEC in THF, PS calibration

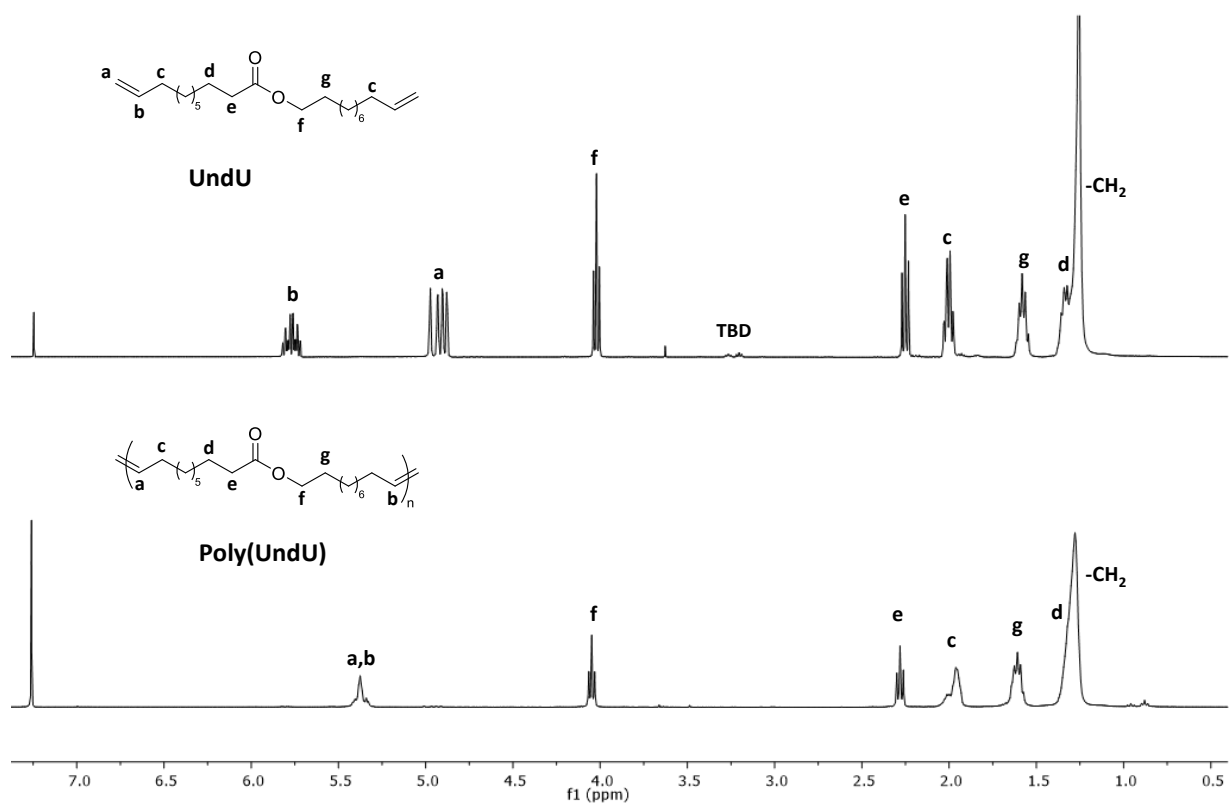

**Figure A1: $^1$H NMR spectra in CDCl$_3$ of UndU and Poly(UndU)**

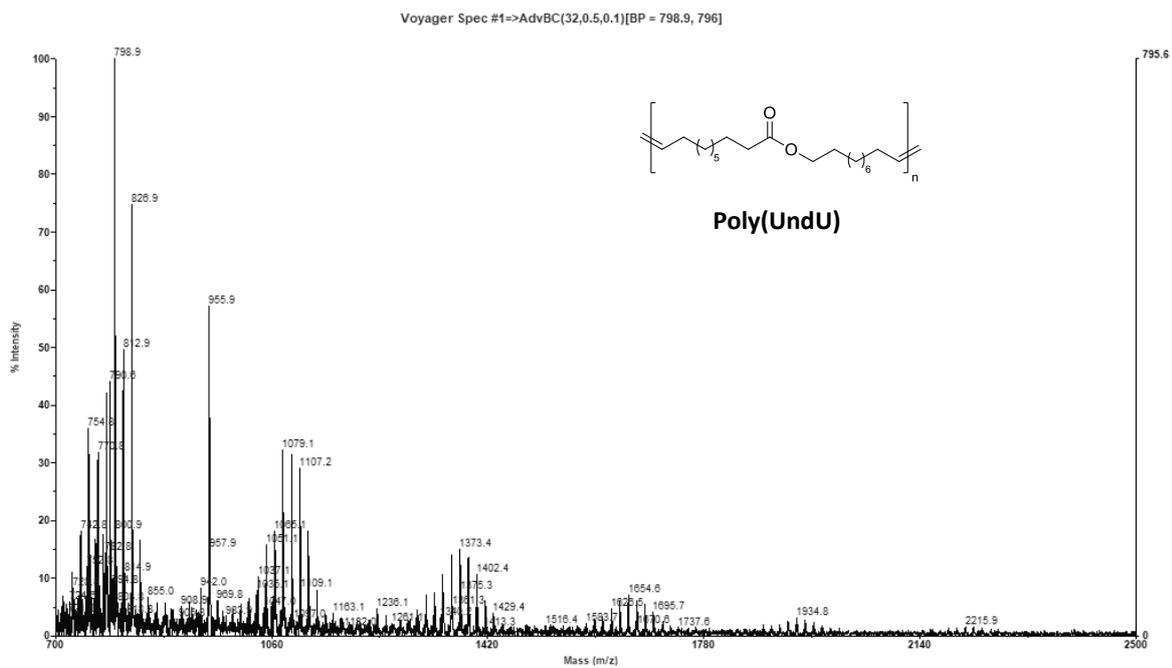

Figure A2: MALDI-TOF of Poly(UndU), matrice DTCB

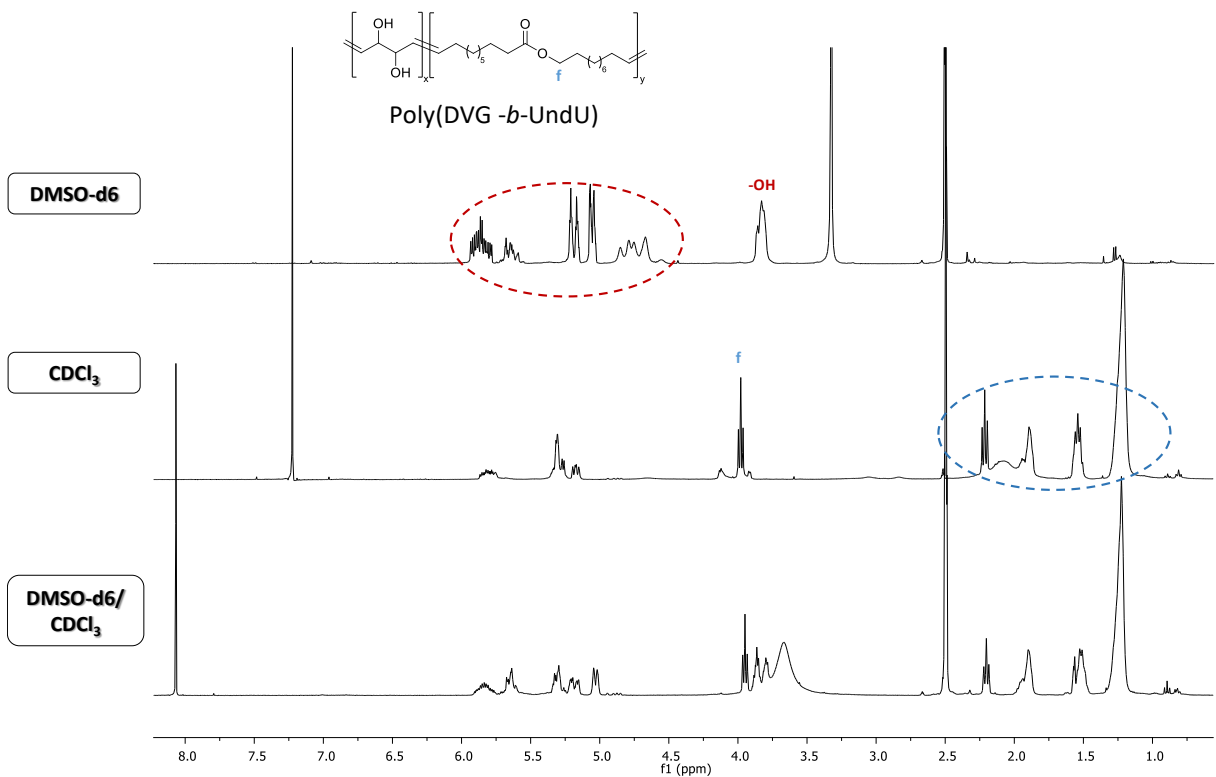

Figure A3: $^1$H NMR spectra of Poly(DVG-UndU) in DMSO, CDCl$_3$ and a mixture of DMSO/CDCl$_3$ (50/50 v/v), respectively

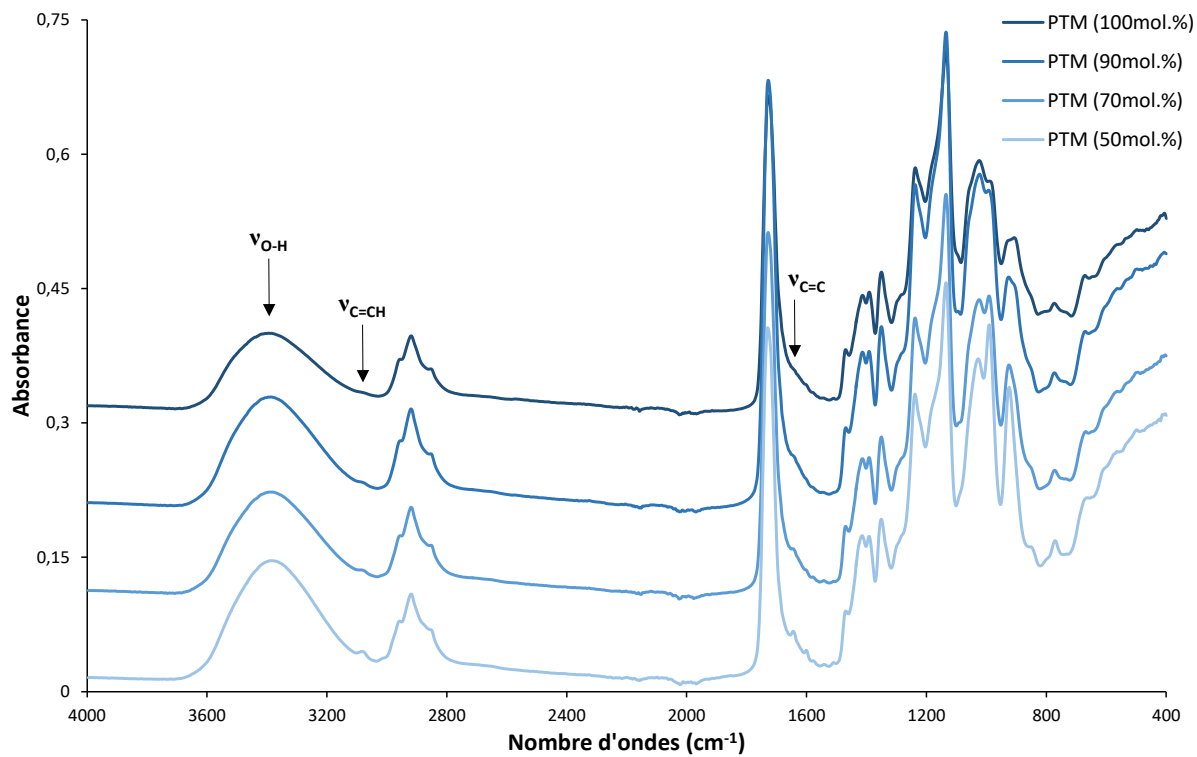

**Figure A4: FTIR spectra of networks synthesized with DVG and PTM**

**Table A3: Characteristics of diamines used to formulated epoxide/amine networks**

| Diamine | f | AHEW | ΔH (J.g$^{-1}$) | T$_{on\,set}$ (°C) | T$_g$ (°C) |
|---|---|---|---|---|---|
| **EDA** | 4 | 15 | 524 | 30 | 86 |
| **PDA** | 4 | 18,5 | 639 | 30 | 69 |
| **DETA** | 5 | 20,63 | 647 | 31 | 73 |
| **IPDA** | 4 | 42,56 | 422 | 39 | 124 |
| **CHMA** | 4 | 35,56 | 597 | 34 | 86 |

**Table A4: Characteristics of networks synthesized with DAG and *Jeffamines***

| Diamine | $T_d^{5\%}$ (%) | $T_g$ (°C) | $F_{max}$ (N) | $W_{adh}$ (N/s) | $t_{rupture}$ (s) | $Gap_{rupt}$ (mm) |
|---|---|---|---|---|---|---|
| Jeffamine ED600 | 287 | -30 | 0,2 | 0,2 | 30 | 30 |
| Jeffamine ED900 | 289 | -42 | 0,35 | 4,1 | 10 | 10,8 |

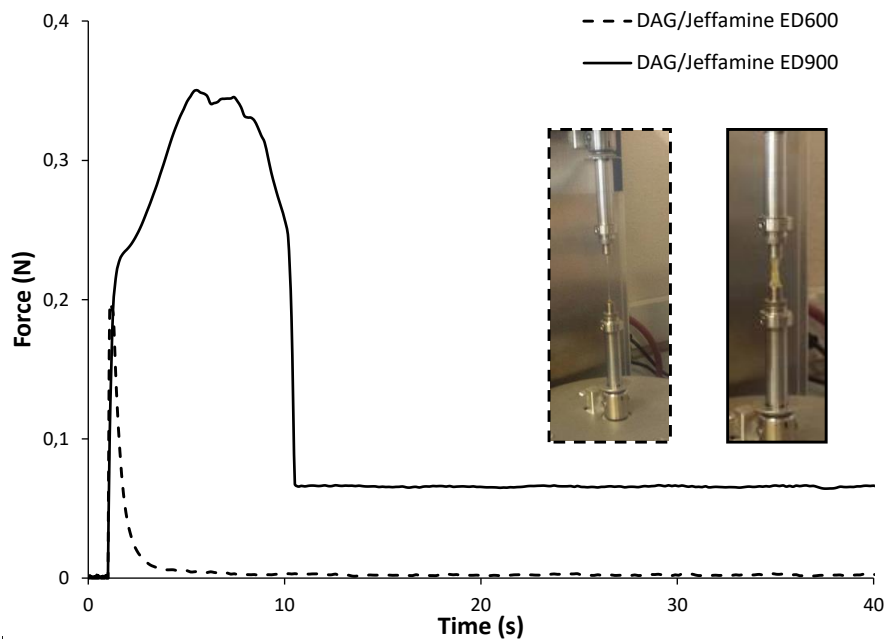

**Figure A5: Tack property analysis of networks synthesized with DAG and *Jeffamine ED600* or *ED900***

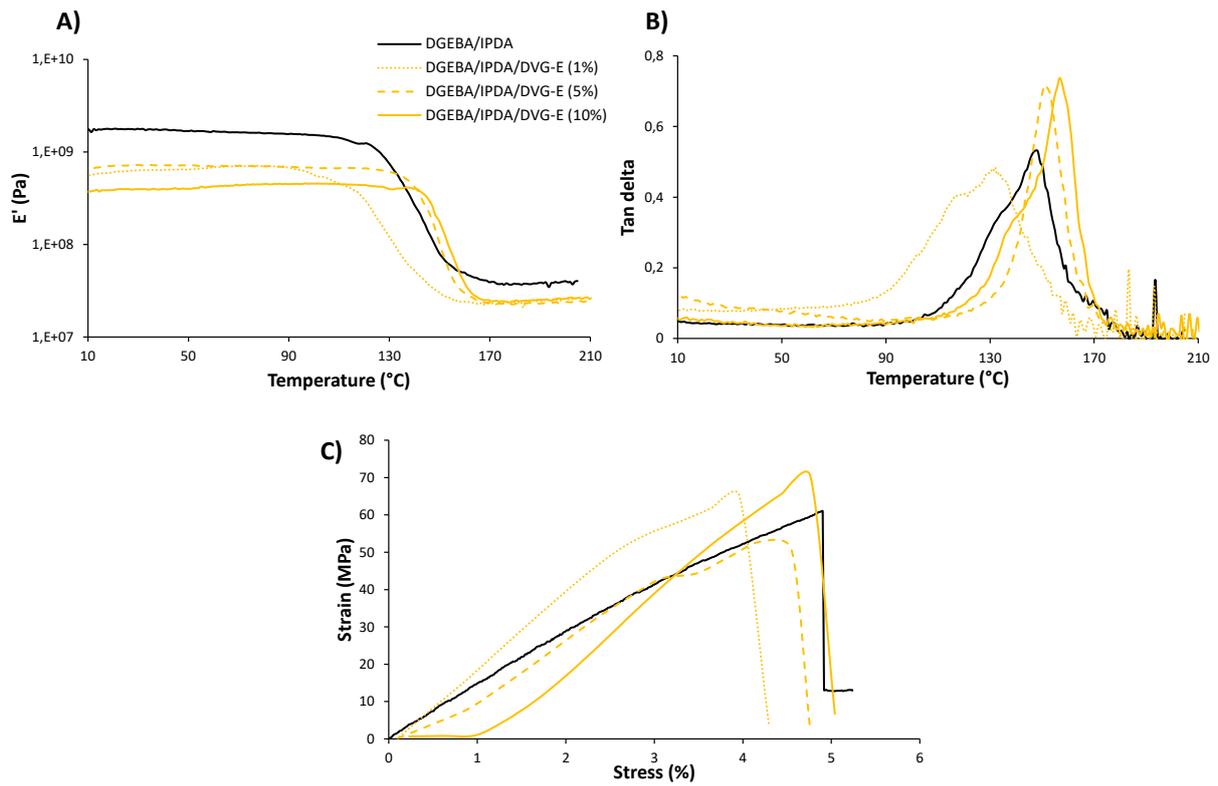

**Figure A6:** Analysis of cross-linked epoxide/amine networks synthesized with DGEBA/DVG and IPDA. A) Young modulus, B) Tan delta, C) tensile tests

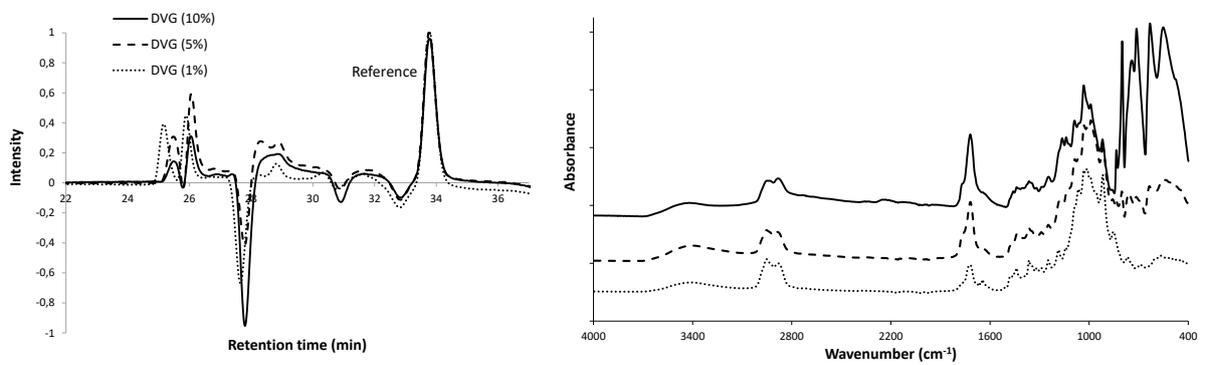

**Figure A7:** Analysis of soluble part after cleavage reaction, SEC in THF, PS calibration (left) and FTIR analysis (right)